\documentclass[aps,prb,twocolumn,showpacs]{revtex4}
\usepackage{mathptm}  
\usepackage{dcolumn}                    
\usepackage{bm}                        
\usepackage{graphicx}
\usepackage{times}
\usepackage{epstopdf}
\usepackage{amssymb}
\usepackage{amsmath}
\usepackage{amsfonts}
\usepackage{color}

\newcommand{\M}{\boldsymbol{M}}

\newcommand{\e}[1]{\mathrm{e}^{#1}}

\newcommand{\eg}{\textit{e.g. }}
\newcommand{\etal}{\emph{et al.}}

\begin{document}

\title{Spin Supercurrent, Magnetization Dynamics, and $\varphi$-State in \\
Spin-Textured Josephson Junctions}

\author{Iryna Kulagina}
\affiliation{Department of Physics, Norwegian University of
Science and Technology, N-7491 Trondheim, Norway}

\author{Jacob Linder}
\affiliation{Department of Physics, Norwegian University of
Science and Technology, N-7491 Trondheim, Norway}

\date{\today}

\begin{abstract}
The prospect of combining the dissipationless nature of superconducting currents with the spin-polarization of magnetic materials is interesting with respect to exploring superconducting analogues of topics in spintronics. In order to accomplish this aim, it is pivotal to understand not only how such spin-supercurrents can be created, but also how they interact dynamically with magnetization textures. In this paper, we investigate the appearance of a spin-supercurrent and the resulting magnetization dynamics in a textured magnetic Josephson current
by using three experimentally relevant models: \textit{i)} a superconductor$\mid$ferromagnet$\mid$superconductor (S$\mid$F$\mid$S) junction with spin-active interfaces, \textit{ii)} a S$\mid$F$_1$$\mid$F$_2$$\mid$F$_3$$\mid$S Josephson junction with a ferromagnetic trilayer, and \textit{iii)} a Josephson junction containing a domain wall. In all of these cases, the supercurrent is spin-polarized and exerts a spin-transfer torque on the ferromagnetic interlayers which causes magnetization dynamics. Using a scattering matrix formalism in the clean limit, we compute the Andreev-bound states and resulting free energy of the system which in turn is used to solve the Landau-Lifshiftz-Gilbert equation. We compute both how the inhomogeneous magnetism influences the phase-dependence of the charge supercurrent as well as the magnetization dynamics caused by the spin-polarization of the supercurrent. Using a realistic experimental parameter set, we find that the spin-supercurrent can induce magnetization switching that is controlled by the superconducting phase difference. Moreover, we demonstrate that the combined effect of chiral spin symmetry breaking of the system as a whole with interface scattering causes the systems above to act as phase batteries that may supply any superconducting phase difference $\varphi$ in the ground state. Such a $\varphi$ junction is accompanied by an anomalous supercurrent appearing even at zero phase difference, and we demonstrate that the flow direction of this current is controlled by the chirality of the magnetization configuration.

\end{abstract}
\maketitle

\section{Introduction}
The synergistic effects of combining ferromagnetism and superconductivity, two seemingly disparate phenomena, have garnered much attention in recent years\cite{buzdin_rmp_05, bergeret_rmp_05}. Investigations regarding the mutual interplay between these condensed phases may be traced back to the early work of Ginzburg \cite{ginzburg} and it is by now established that ferromagnetic order not necessarily acts detrimentally toward superconductivity - the two may even coexist in a series of uranium-based heavy fermion compounds such as UGe$_2$, UCoGe, and UIr \cite{saxena, aoki, huy_prl_07}. Whereas such systems pose several challenges with regard to experimental investigations \eg due to requirements of very high pressures in some cases, the combined influence of FM and SC order can be studied in a more controllable fashion by tailoring hybrid structures with the desired properties. 

The physical mechanism behind the unlikely alliance of magnetic and superconducting order is symmetry breaking combined with the Pauli exclusion principle \cite{eschrig_jtlp_07}. As long as the Cooper pair wavefunction respects the correct antisymmetry property under an exchange of the particle-coordinates for spin, space, and time, the Cooper pairs can in fact become spin-polarized. Such an effect takes place in FM/SC structures since both the explicit translation symmetry breaking due to the interface and the presence of a band-splitting exchange field creates Cooper pairs with different symmetry properties than in the bulk superconductor \cite{tanaka_prl_07}. The consequence of same-spin electrons constituting a Cooper pair is that they become insensitive to the paramagnetic limitation of internal or external magnetic fields, allowing such correlations to survive distances up to hundreds of nanometer inside a ferromagnet \cite{bergeret_prl_01}, even in extreme cases such as half-metallic compounds \cite{eschrig_prl_03, keizer}. In such a scenario, the limiting factor of the penetration depth is not determined by the strength of the magnetic exchange field, but by other pair-breaking events such as spin-flip and inelastic scattering \cite{birge_prb_11}. Experiments have unambiguously observed such long-ranged superconducting correlations arising in FM/SC structures that feature magnetic textures of some sort: this includes multilayered magnetic structures \cite{robinson_science_10, khaire_prl_10}, domain wall or intrinsically textured ferromagnets \cite{sosnin_prl_06, robinson_scirep_12}, and interfaces with spin-active scattering and/or disorder \cite{anwar_prb_10, sprungmann_prb_10}. A large amount of theoretical work has recently been devoted to the topic of spin-triplet correlations arising in S/F hybrid structures (see \eg \cite{pajovic_prb_06, houzet_prb_07, halterman_prl_07, eschrig_natphys_08, brydon_jpsj_08, alidoust_prb_10, baker_arxiv_13, shomali_njp_11, cottet_prl_11, buzdin_prb_11, sperstad_prb_08, fominov_prb_07, annunziata_prb, halterman_prb_08, romeo_prl_13, grein_prl_09, asano_prb_07, yokoyama_prb_07, lindercuoco_prb_10, halasz_prb_09, margaris_jpcm_10, bergeret_prb_12, liu_prb_10}).

The existence of long-ranged spin-polarized superconducting correlations raises an interesting question: is it possible to utilize this to obtain a superconducting analogue to central topics in spintronics such as domain wall motion  and magnetization switching? It is well-known that resisitive (normal) spin-polarized currents play a central part in terms of obtaining magnetization dynamics in spintronics \cite{zutic_rmp_05}. Spin-currents enable a transfer of angular momentum to the magnetic order parameter of a material via the effect of spin-transfer torque \cite{slon, berger}. Since spin-supercurrents also carry angular momentum, the same effect is possible in this context and a few previous works have investigated the possibility of magnetization dynamics in superconducting hybrid structures \cite{waintal_prb_02, zhao_prb_07, buzdin_prl_08, braude_prl_08, linder_prb_11, holmqvist_prb_11, sacramento}. However, it remains unclear how the superconducting phase difference affects the dynamics via the Andreev bound-state spectrum. In this paper, we will consider three experimentally relevant types of FM/SC weak-link structures that all have in common that the region separating the superconductors is spin-textured. We will compute the spin-polarized supercurrent analytically, and demonstrate that its spin-torque can give rise to magnetization switching by solving the non-linear Landau-Lifshitz-Gilbert \cite{LLG} equation numerically. This constitutes a way to directly utilize the spin-polarized nature of the recently observed long-range triplet currents in order to dynamically alter magnetization textures. In addition to this, we will demonstrate that the magnetic structure in such Josephson junctions has a profound effect on the superconducting ground-state itself. Whereas it is known that superconductor$\mid$ferromagnet$\mid$superconductor (S$\mid$F$\mid$S) junctions normally have a ground-state phase difference of 0 or $\pi$, it was very recently demonstrated experimentally that it is possible to construct a $\varphi$-state junction where the ground-state phase takes on any value between 0 and $\pi$ \cite{sickinger_prl_12}. Such a $\varphi$-state was originally proposed to occur in SFS junctions in \cite{buzdin_prb_05} and subsequently studied in several works 
\cite{braude_prl_07, buzdin_prl_08, linder_prb_10, liu_prb_10_soc, goldobin_prl_12, alidoust_prb_13, heim_jpcm_13, feinberg_arxiv_14}, offering the unique possibility to design phase batteries \cite{ortlepp_science_06, feofanov_natphys_10} with an arbitrary phase-shift rather than only 0 or $\pi$  which could be used to bias both classical and quantum circuits.  We will compute the free-energy and belonging supercurrent-phase relation in inhomogeneous magnetic Josephson junctions and show that anomalous behavior arises in the form of a finite supercurrent even at zero phase difference. As will be shown, this is intimately linked with a chiral spin symmetry breaking and scattering at the interfaces of the structure and results in the possibility of a controllable $\varphi$-state by adjusting the magnetization vectors in the system. 

\begin{figure}[t!]
\centering
\resizebox{0.48\textwidth}{!}{
\includegraphics{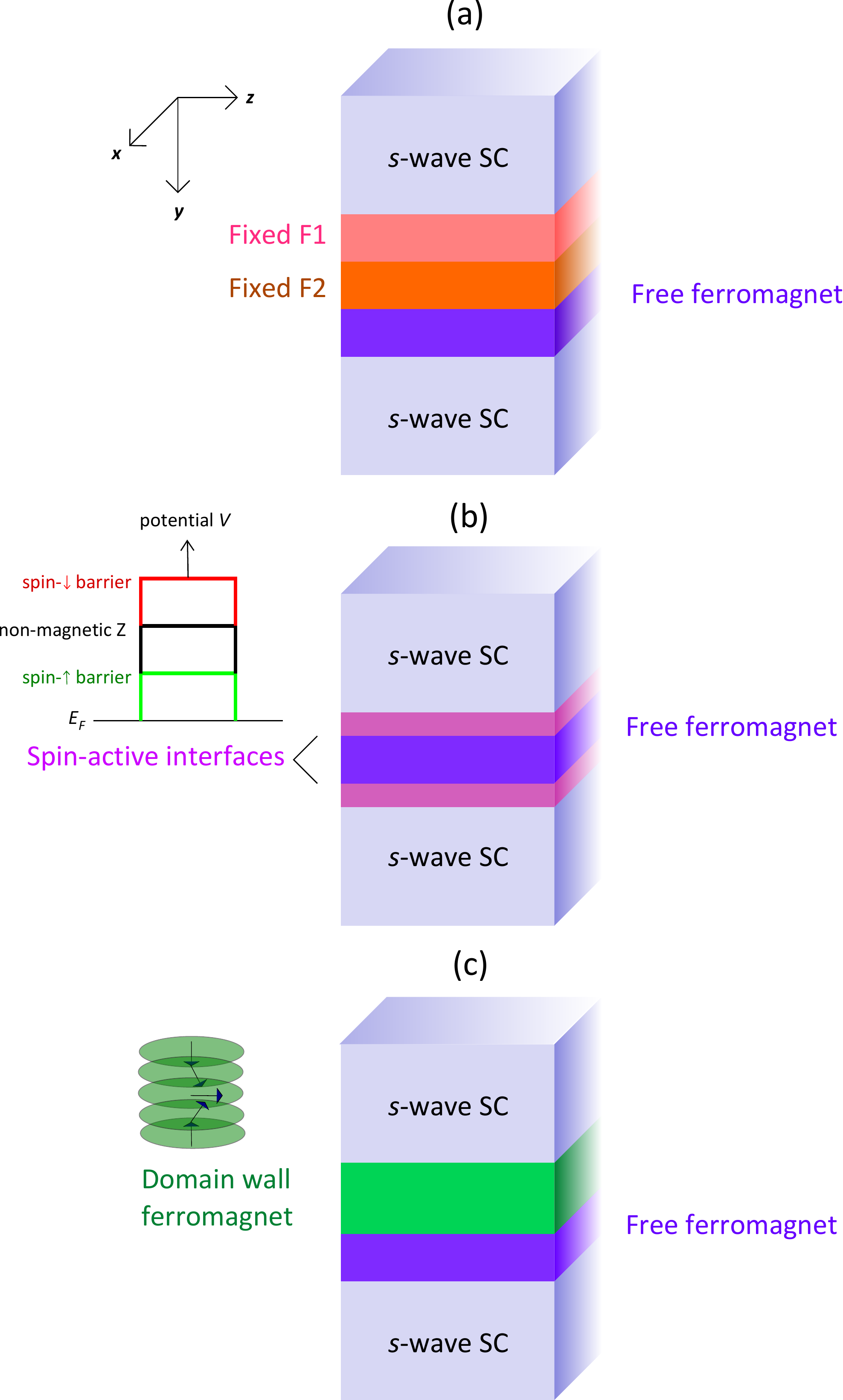}}
\caption{(Color online) The three setups considered in this paper for magnetization dynamics induced by a spin-polarized supercurrent: (a) a trilayer S$\mid$F$\mid$S junction with non-collinear magnetization, (b) S$\mid$F$\mid$S junction with spin-active interfaces, and (c) S$\mid$DW$\mid$F$\mid$S junction where the supercurrent is polarized by a domain-wall. }
\label{fig:model} 
\end{figure}

This paper is organized as follows. In Sec. \ref{sec:theory}, we outline the theoretical framework used in our calculations of the spin supercurrent, Andreev levels, the magnetization dynamics, and the ground-state energy of the system under consideration (see Fig. \ref{fig:model}). In essence, we are combining the mean-field Bogolioubov-de Gennes equations in a scattering state framework to compute the free energy from which all thermodynamic quantities may be obtained, and then extract the effective magnetic field in our theory which is used as input in the Landau-Lifshitz-Gilbert equation in order to obtain the magnetization dynamics. Additional details of the calculations are found in the Appendix. In Sec. \ref{sec:results}, we give a comprehensive treatment of the Andreev levels that arise and compute the spin-polarized supercurrent flowing in the system. We provide results for the current-phase relation and magnetization dynamics, and show how a $\varphi$-state may arise in non-collinear arrangements in addition to magnetization switching. We give a detailed discussion of our results in Sec. \ref{sec:discussion}, in particular with regard to the experimental feasibility of our proposed setup and the regime of validity for the approximations made in our calculations. Finally, we summarize our findings in Sec. \ref{sec:summary}.

\section{Theory}\label{sec:theory}
We consider a ballistic Josephson junction composed of one or more ferromagnetic layers sandwiched between two conventional s-wave superconducting electrodes. The entire structures is positioned along the $y$-axis such that the interfaces lie in the $x-z$-plane. We choose the origin $y=0$ to be at the interface between the left superconducting layer and its proximate ferromagnetic. Assuming large superconducting banks with size $d\gg\xi_S$, these layers are characterized by their bulk superconducting gap $\Delta$ and the macroscopic phase difference across the junction, $\gamma=\gamma_{R}-\gamma_{L}$.

The ferromagnetic part of the junction depends on the specific model considered as shown in Fig. \ref{fig:model}. We will treat three experimentally relevant model systems in order to illustrate the rich physics that arises due to the spin-polarized nature of the long-ranged superconducting correlations. In Fig. \ref{fig:model}(a), we consider a multilayered ferromagnetic junction, similar to a recent experiment \cite{khaire_prl_10}. As predicted by Ref. \cite{houzet_prb_07}, the Josephson current in such a structure should have a long-ranged contribution that depends on the relative orientation of the magnetization vectors in each of the ferromagnetic layers. To treat a general scenario, we consider an arbitrary direction of the magnetization in the free layer and fix the orientation in the two hard magnetic layers to the $z$- and $x$-axis, respectively. The three layers $j\in\{1,2,3\}$ are characterized by their thickness $L_j$ and exchange field $h_j$, and we will also consider the influence of interface resistance captured by an effective dimensionless parameter $Z$ (see Appendix). As we will calculate below, the rich physics including supercurrent-induced magnetization reversal and the appearance of a $\varphi$-ground state is intimately related to chiral symmetry breaking by the magnetization vectors $\boldsymbol{M}_j$ \cite{asano_prb_07, margaris_jpcm_10}, characterized by a finite value of the chirality vector:
\begin{align}
\boldsymbol{\chi} = \boldsymbol{M}_1 \cdot (\boldsymbol{M}_2\times\boldsymbol{M}_3).
\end{align}

Next, we consider in Fig. \ref{fig:model}(b) a free magnetic layer with low anisotropy where the interface region coupling to the superconductors is spin-active. Such interfaces are known to give rise to spin mixing, and spin rotation \cite{eschrig_natphys_08} which considerably alters the superconducting proximity effect. We consider a situation where the barrier moments lie in the $x-z$-plane with the parallel, perpendicular and antiparallel alignments given by $\phi_L=\phi_R=0$, $\phi_L=0$ and $\phi_R=\pi/2$, $\phi_L=0$ and $\phi_R=\pi$, respectively. The spin-active interfaces are characterised by barriers \cite{lindercuoco_prb_10}
\begin{align}\label{eq:sa}
U=[\hat{1}-\rho_m\cos(\phi)(\tau_0\otimes\sigma_3)-\rho_m \sin(\phi)(\tau_0\otimes\sigma_1)],
\end{align}
One of our results is that breaking chiral spin symmetry is not a sufficient condition to generate an anomalous zero-phase difference supercurrent. Instead, the scattering taking place at the interfaces separating the various regions will be shown to play a pivotal part in this. Finally, we include the effect of a domain wall by considering in Fig. \ref{fig:model}(c) a setup where the ferromagnetic region consists of a domain wall and a free magnetic layer. The domain wall is taken to be of Bloch-type, thus rotating around the $y$-axis with a characteristic length scale of $\lambda$. This particular choice of domain wall is not essential to the resulting physics, and the results we obtain are qualitatively unchanged for other types of magnetization textures. The structure of the domain wall is described by a vector $\boldsymbol{f}$ proportional to the magnetization vector \cite{walker}. In order to obtain analytical results, we use the following form:
\begin{equation}\label{eq:dwprofile}
  \boldsymbol{f}(y)=\begin{cases}
    [\sin(\frac{\pi y}{l_\text{dw}}),  0,  \cos(\frac{\pi y}{l_\text{dw}})], & \text{if $0<y<l_\text{dw}$}.\\
    0, & \text{otherwise}.
  \end{cases}
\end{equation}
The starting point for all scenarios described above is the mean-field Bogoliubov-de Gennes equations \cite{BDG} describing quasiparticle propagation in these structures. Due to the non-collinear magnetization textures, one must consider the full spin$\otimes$particle-hole space and use a four-component wave function $\Psi=(u_\uparrow, u_\downarrow, v_\uparrow, v_\downarrow)^T$
\begin{equation}
\begin{pmatrix}
\hat{H_0}(y) & \hat{\Delta}(y)\\ 
-\hat{\Delta}^\dagger (y) & -\hat{H_0}^T (y)
\end{pmatrix} \Psi(y) = E\Psi(y)
\end{equation}
where $\hat{\Delta}(y)=i\sigma_2 \Delta(y)$ and the single-particle Hamiltonian is
\begin{equation}
\hat{H_0}(x)=\left [ -\frac{\bigtriangledown}{2m}-\mu(y) \right ]\hat{1}-h \boldsymbol{f}(y)\cdot\boldsymbol{\sigma}
\end{equation}
where $m$ is effective mass of quasiparticles, $\mu$ is chemical potential, and $\boldsymbol{\sigma}$ is the Pauli matrix spin-vector. The quasiparticle energy $E$ is measured relative the chemical potential which in the low-temperature limit considered here equals the Fermi energy. The eigenstates $\Psi$ may be constructed once the magnetization texture $\boldsymbol{f}(y)$ is specified [see Fig. \ref{fig:model}]. In each case, the free layer magnetization is allowed to take arbitrary directions. This enables a study of the supercurrent-induced magnetization dynamics on the magnetic order parameter of this layer. We also mention that the scattering states in the domain wall region treated in case (c) may be obtained by employing a unitary transformation of the Hamiltonian which rotates the spin-basis to follow the magnetization texture. This also alters the boundary conditions to the superconducting regions. All of these calculational details are left for the Appendix. 

Using the framework sketched above, one may compute the allowed energy-levels that exist in the Josephson junctions. These Andreev levels $\varepsilon$ will depend on the junction geometry, the $U(1)$ superconducting phase gradient, and the magnetization texture. With them in hand, both the free energy $\mathcal{F}$ and the charge supercurrent $\mathcal{I}$ are obtained via \cite{beenakker_prl_91}:
\begin{align}\label{eq:FandI}
\mathcal{F}(\gamma) = -\frac{1}{\beta} \sum_j \text{ln}(1 + \e{-\beta \varepsilon_j}),\; \mathcal{I}(\gamma) = \frac{2e}{\hbar} \sum\limits_{i}f(\varepsilon_i)\frac{\partial \varepsilon_i}{\partial \gamma}
\end{align}
where $f(\varepsilon)$ is Fermi-Dirac distribution function and $\beta=1/k_BT$. The fact that the supercurrent is spin-polarized due to the long-range triplet proximity effect and flows under equilibrium conditions directly implies that the exchange interaction between the ferromagnets will be altered by the superconducting phase difference $\gamma$. In fact, there is an interesting co-dependence between the phase difference $\gamma$ and the non-collinearity of the magnetization vectors regarding the supercurrent $\mathcal{I}$ and the equilibrium magnetic torque $\tau$ as first noted by Waintal and Brouwer \cite{waintal_prb_02}. Considering for simplicity two monodomain ferromagnets with a relative angle $\theta$ between the magnetization vectors, it follows from $\mathcal{I} = \frac{2e}{\hbar}\frac{\partial \mathcal{F}}{\partial \gamma}$ and $\mathcal{\tau} = \frac{\partial \mathcal{F}}{\partial \theta}$ that:
\begin{align}
\frac{\partial\mathcal{I}}{\partial\theta} = \frac{2e}{\hbar}\frac{\partial\tau}{\partial\gamma}.
\end{align}
The above equation is simple, yet it conveys a powerful message: if the supercurrent is sensitive to the magnetization orientation, then the torque exerted on the magnetic order parameters is sensitive to the superconducting phase difference. This is the core principle which enables the supercurrent-induced magnetization dynamics in inhomogeneous S$\mid$F$\mid$S junctions. The induced superconducting correlations are long-ranged since they become spin-polarized and thus avoid picking up a finite center-of-mass momentum which acts pair-breaking. In turn, their spin-polarized nature makes them sensitive to the magnetization texture in the junction such that a mutual interplay is enabled between the supercurrent and the magnetization. 

Having obtained the free energy of the system from the Andreev levels, one may also compute the effective field $\boldsymbol{H}_\text{eff}$ that couples to the magnetic order parameter:
\begin{align}
\boldsymbol{H}_\text{eff} = -\frac{1}{V}\frac{\partial\mathcal{F}}{\partial \boldsymbol{M}}
\end{align}
The effective field is used to describe the supercurrent-induced magnetization dynamics in the free layer (blue region in Fig. \ref{fig:model}) by solving the Landau-Lifshitz-Gilbert equation \cite{LLG}:
\begin{align}
\frac{\partial \M}{\partial t} = -\zeta \M \times \boldsymbol{H}_\text{eff} + \alpha \M \times \frac{\partial \M}{\partial t},
\end{align}
where $\zeta$ is the gyromagnetic ratio and $\alpha$ is the Gilbert damping constant. As long as the effective field is not fully aligned with the magnetization, it will exert a torque on it which induces magnetization dynamics. We are considering a monodomain macrospin model for the soft ferromagnetic layer, such that there is no contribution from the spin stiffness term $\sim \frac{\partial^2\M}{\partial y^2}$. However, we include the influence of magnetic anisotropy with additional terms $\pm K_j M_j^2$, $j\in\{x,y,z\}$ in the free energy where $K_j$ are the anistropy constants and the $\pm$ sign determines the hard and easy axes of magnetization.

\section{Results}\label{sec:results}

We will now proceed to present our results for the Andreev bound-state (ABS) spectrum, the system's free energy, the current-phase relation, and the ensuing magnetization dynamics via spin-supercurrents. We treat each of the three proposed systems in Fig. \ref{fig:model} separately. In each subsection, we start by considering the analytical expression for the ABS energy. Obtaining this quantity serves as the foundation for the computation of both the total free energy of the system and the equilibrium supercurrent, as given by Eq. (\ref{eq:FandI}). The technical procedure for doing so consists of three steps. First, we obtain the eigenstate wavefunctions that solve the BdG equations in each region (see Appendix for details). From these wavefunctions, the appropriate scattering states involving particle- and hole-like excitations are constructed with belonging probability coefficients. The energies $\varepsilon$ that allow for a non-trivial solution of the scattering coefficients are obtained by matching the wavefunctions at each interface region using appropriate boundary conditions and setting up a system of linear equations of the type $\hat{A}\mathbf{x} = \mathbf{b}$ where $\mathbf{x}$ contains the scattering coefficients. Solving the characteristic equation det$\hat{A}=0$ allows one to identify the ABS solutions for $\varepsilon$. The boundary conditions require some special care for the systems under consideration in the present paper, i.e. they are modified from conventional boundary conditions both for setup (b) and (c) in Fig. \ref{fig:model}. 

\subsection{Trilayered S$\mid$F$\mid$F$\mid$F$\mid$S structure}
The magnetizations in the first two layers F$_1$ and $F_2$ are assumed to be fixed via strong anisotropy energies along the $\hat{z}$ and $\hat{x}$ directions, respectively. In F$_3$, we allow for an arbitrary magnetization direction in order to explore the effect of spin-supercurrent induced magnetization dynamics. This material should then consist of a much softer ferromagnet than F$_1$ and F$_2$. For a completely arbitrary parameter set, the analytical expression for the ABS-energy is overwhelming. However, physical insight can be obtained in experimentally relevant limiting cases. In the quasiclassical regime of a rather weak ferromagnet $h/\mu\ll1$, one finds that:
\begin{align}\label{eq:energytrilayer}
\varepsilon_j = \Delta_0\sqrt{1 - \mathcal{A}\cos\gamma+\mathcal{B} Z^3 (h_y/h) \sin\gamma - \mathcal{C} \pm \sqrt{\mathcal{D} (\gamma)}}
\end{align}
where the coefficients $\mathcal{A}, \mathcal{B}, \mathcal{C}$ are  independent on the phase difference $\gamma$. Instead, they are functions of the junction parameters such as length $L$, barrier $Z$, and exchange field $h$. It should be noted that Eq. (\ref{eq:energytrilayer}) is valid for arbitrary interface transparency $Z$. We provide some additional details for the coefficients in Eq. (\ref{eq:energytrilayer}) in the Appendix. The quantity $\mathcal{D}(\gamma)$ is a rather large expression which depends on $\gamma$; the essential property of this quantity is nevertheless that 
\begin{align}
\frac{\partial \mathcal{D}(\gamma)}{\partial \gamma}\Bigg|_{\gamma=0} \propto \mathcal{B} Z^3 (h_y/h).
\end{align}
We prove now that it follows from the above properties of the Andreev-level that there will be a finite supercurrent at zero phase difference. This finding is then independent on the specific details of the coefficients introduced above.

The presence of an anomalous current is seen to be contigent on two factors: 1) the presence of scattering barriers and 2) $h_y\neq0$ in the free F layer. The absence of either of these causes the supercurrent to revert to conventional behavior. We comment first on the role of the scattering barriers. In Eq. (\ref{eq:energytrilayer}), it was assumed that the scattering barrier $Z$ was the same for the interfaces between the ferromagnetic regions whereas the S/F interface was taken to be completely transparent. By allowing for different barrier values, which will be the case in general since the value of $Z$ depends on the specific materials connected, one finds that the term providing the anomalous current reads $\frac{1}{2}\mathcal{B} Z_1Z_2(Z_1+Z_2) h_y \sin\gamma$. Here, $Z_1$ is the barrier between the $F_1/F_2$ interface whereas $Z_2$ is the barrier between the $F_2/F_3$ interface. This demonstrates that in the short-junction regime where the Andreev bound-states carry the current, barriers at both ferromagnetic interfaces are required in order to produce the anomalous current: setting either $Z_1$ or $Z_2$ to zero cancels the $\sin\gamma$ term in Eq. (\ref{eq:energytrilayer}). We will later establish a connection between this observation and the results for the domain wall junction to be considered in a section below.  

Secondly, the fact that the anomalous supercurrent only appears when $h_y\neq0$ means that the presence of an explicitly broken chiral spin symmetry the system is a necessary criterium. Interestingly, we find that direction of the current is actually controlled by the specific chirality, i.e. the sign of $h_y$. A consequence of this is that the magnetization direction then acts as a 0-$\pi$ switch as it controls the direction of the supercurrent, which offers a novel way of exerting dynamical control over a superflow of spins. The precise quantitative behaviour of the system depends also on the following parameters: the interface barrier, the magnetic anisotropy constant, and the length of ferromagnetic layers. For convenience, we introduce the normalized and dimensionless variables $\beta_i=\frac{k_F L_i h}{2\mu}$, where the index $i$ denotes the ferromagnetic layer under consideration. Throughout this work, we set $k_F L=2\pi n$, where $n$ is integer. The presence of ferromagnetism introduces additional phase-shifts for the Andreev bound-states as they propagate through the system.

\begin{figure}[t!]
\centering
\resizebox{0.48\textwidth}{!}{
\includegraphics{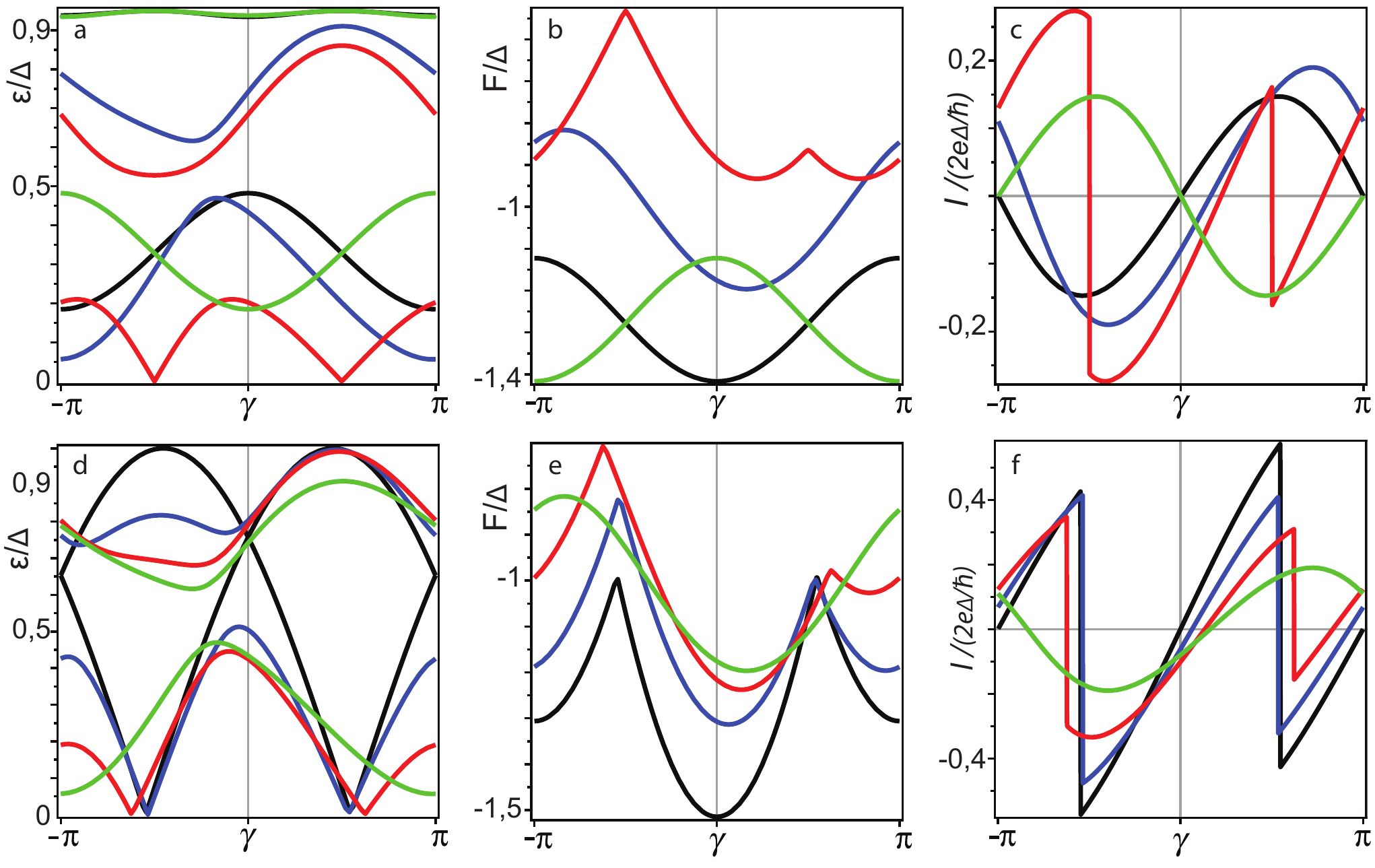}}
\caption{(Color online) (a,d): Andreev bound-state energies as a function of superconducting phase difference $\gamma$. (b,e): free energy of the system as a function of $\gamma$ and (c,f) supercurrent-phase relation for our trilayered S$\mid$F$\mid$F$\mid$F$\mid$S structure. In all plots, we have set $\beta_1=\beta_2=\pi/3$. In (a,b,c), we fix the barrier at $Z=2$ and investigate the effect of different values of $\beta_3$ (proportional to both exchange field $h$ and width $L$ of the free ferromagnetic layer): $\beta_3=0$ (black), $15\pi/100$ (blue), $25\pi/100$ (red), $50\pi/100$ (green). For (d,e,f), we fix $\beta_3=15\pi/100$ and investigate the effect of a varying barrier potential: $Z=0$ (black), $1$ (blue), $1.5$ (red), $2$ (green).}
\label{fig:energy for SFFFS} . 
\end{figure}

In Fig. \ref{fig:energy for SFFFS}, we plot the ABS-energy (a, d), the free energy (b, e) and the Josephson current (c, f) as function of the phase difference. We fixed $\beta_1=\beta_2=\pi/3$ and considered several values for $\beta_3$ and $Z$. The magnetization in the free layer has been set to $\mathbf{m} \parallel \hat{y}$ in order to demonstrate the appearance and consequences of the anomalous supercurrent. To give the reader a better idea about which values these correspond to in an experimental setup, we note that for a weakly polarized ferromagnet with $h/\mu=0.02$ (exchange field of around 30 meV), $\beta=\pi/3$ corresponds to a length of 15 nm. In Fig. \ref{fig:energy for SFFFS}, we consider in (a-c) the effect of varying the width or exchange field of the free ferromagnetic layer, captured in the parameter $\beta_3$. We consider here a weakly transparent interface $Z=2$. In (d-f), we instead fix $\beta_3$ and consider the influence of having different barrier potentials $Z$. The panels for the ABS-energies clearly display that the current is spin-polarized as their spin-degeneracy is completely removed in the present system. One important feature is that the effect of increasing $Z$ on the spectrum is that the maxima and minima are shifted away from a phase difference $\gamma=0$ and $\gamma=\pi$. The fact that the derivative of the ABS-energy with respect to $\gamma$ does not vanish at these points implies that there will be a finite current even in the absence of any superconducting phase difference. This will be referred to as an \textit{anomalous supercurrent}. We observe that there is no anomalous supercurrent when $Z=0$, as seen also in Eq. (\ref{eq:energytrilayer}).

The presence of an anomalous supercurrent is intimately related to an unusual property for the quantum ground-state of the system, which is illustrated in the plots for the free energy in Fig. \ref{fig:energy for SFFFS} (b) and (e). The global minimum of $F$ is seen to not necessarily occur at the conventional 0 and $\pi$ states for the phase difference - in fact, for weakly transparent interfaces it deviates strongly from these values and occurs at an intermediate phase $\in[0,\pi]$. This is a manifestation of a so-called $\phi$-junction. In the right column of Fig. \ref{fig:energy for SFFFS}, we plot the supercurrent-phase relation for various choices of the length and exchange field for the free ferromagnetic layer as well as different values of the interface transparency. When a $\phi$-junction is realized, we have $I(\gamma=0)\neq0$ and an anomalous current is present. Its magnitude is strongly dependent on $\beta_3 \propto hL$ and $Z$, and is seen to reach up to 50\% of the critical Josephson current (for $\beta_3=\pi/4$ in the figure under consideration). 

Having considered the equilibrium properties of the magnetically textured trilayer-Josephson junction, we now wish to address if magnetization dynamics will be generated when a spin-polarized supercurrent flowing through the system. In particular, we will consider if and how the presence of the aforementioned anomalous supercurrent alters the dynamics of the free ferromagnetic layer. To explore this, we solve the Landau-Lifshitz-Gilbert (LLG) equation numerically without any approximation for the ABS energies, i.e. valid for arbitrary parameter values. The main ingredient which makes this possible is the effective field, which contains both the contribution from anisotropy terms and the ABS-energies. It may be written as:
\begin{align}
\textbf{H}_\text{eff} = \frac{2}{|M_0|} (K_{e} \boldsymbol{m}_i-K_{h} \boldsymbol{m}_j)-\frac{1}{V |M_0|} \frac{\partial \mathcal{F}}{\partial \boldsymbol{m}}
\end{align}
where $K_{e(h)}$ is the easy (hard) axis anisotropy constant while $\mathcal{F}$ is the contribution to the free energy from the ABS-energies [see Eq. (\ref{eq:FandI})] and $i (j)$ can be $x$ or $y$ or $z$ in accordance with in which direction is easy (hard) axis. 
We comment specifically on the regime of validity for our approach that consists of combining a scattering matrix approach in equilibrium with the time-dependent LLG-equation in Sec. \ref{sec:discussion}. For now, we simply state that this framework is justified when the magnetization dynamics is sufficiently slow compared to the rate at which the system relaxes to an equilibrium state \cite{tserkovnyak_rmp_05}, and is commonly used in the literature. In our numerical simulations, we will set $\beta_1=\beta_2=\pi/3$, $\Delta=10^{-22}$ J, $\mu_0=10^{-6}$ H/m, and $|M_0|=10^{5}$ A/m. The Gilbert damping parameter is set to $\alpha=0.02$. 

\begin{figure}[t!]
\centering
\resizebox{0.50\textwidth}{!}{
\includegraphics{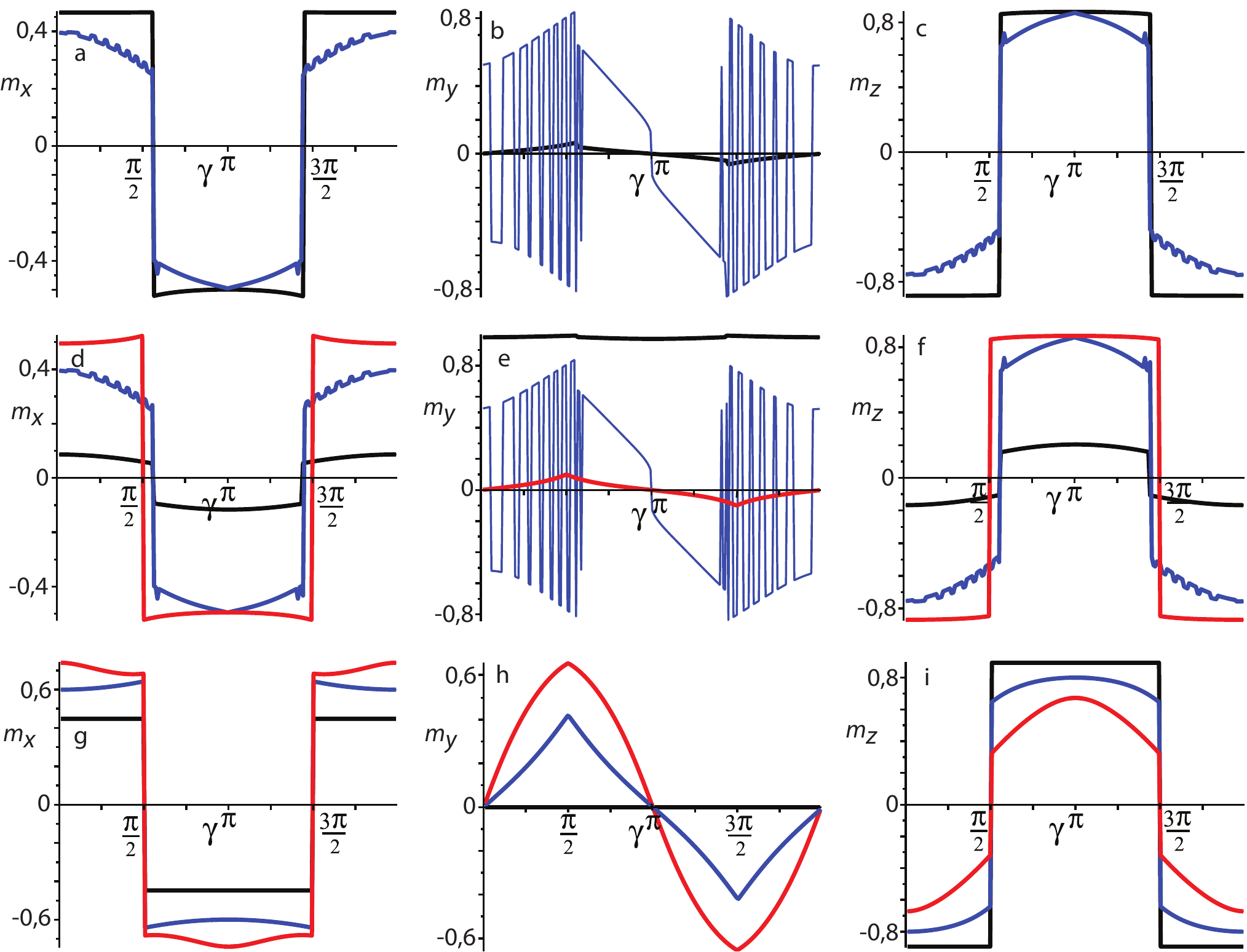}}
\caption{(Color online) Stable magnetization state as a function of superconducting phase difference $\gamma$ for $t\to\infty$ when $\textbf{m}_3(t=0) \parallel \hat{y}$ initially. The components of the magnetization are given in the left ($m_x$), middle ($m_y$), and right $(m_z)$ columns. For all panels, we fix $\beta_1=\beta_2=\pi/3$. (a,b,c): We set $\beta_3=5\pi/100$, $Z=0.5$, and consider different values of the anisotropy constant - $K=10^4$ J/m$^3$ (black line), $10^5$ (blue line). (d,e,f): We set $Z=0.5$, $K=10^5$, and consider different values of the $\beta_3$ parameter - $\beta_3=\pi/100$ (black), $5\pi/100$ (blue), $25 \pi/100$ (red). (g,h,i): We set $\beta_3=25\pi/100$, $K=10^5$, and consider different values of the barrier transparency - $Z=0$ (black), $1$ (blue), $2$ (red). }
\label{fig:SFFFSmagdyn_y} 
\end{figure}

Before discussing the obtained results, it should be noted that the time-dynamics of the magnetic order parameter in the free F layer depends on the relative magnitude of the anisotropy and ABS-energy terms in the effective field $\textbf{H}_\text{eff}$. Depending on the parameters of the system, one of these will dominate or they will be of similar magnitude and compete. We will take the cross-sectional area of the junction to be 1$\mu$m$\times$1$\mu$m and consider a width of 10 nm for the free layer. With a lattice constant of $a=0.1$ nm and estimating the number of transverse modes to $N/V = 10^{28}$ m$^{-3}$, we find that for $K\leq 10^3$ J/m$^3$ the ABS-term dominates whereas for $K\geq10^5$ J/m$^3$ the anisotropy governs the dynamics. In order to limit the parameter space, we will consider only a high to moderate interface transparency ($Z\leq2$) and a junction length of the free F layer satisfying $\beta_3\leq25\pi/100$. These values are representative for a set of experimentally attainable interface transparencies ranging from high to low as well as different values for the exchange field of the free ferromagnetic layer, ranging from weakly to moderately polarized. In each case, we solve the LLG-equation numerically and identify the stable state that arises when $t\to\infty$ and its dependence on the superconducting phase difference. The initial condition for the magnetization of the free layer is taken to be along its easy anisotropy axis. We discuss the experimental realization of this setup in more detail in Sec. \ref{sec:discussion}.

Firstly, consider the case with anisotropy along the $\hat{y}$ direction shown in Fig. \ref{fig:SFFFSmagdyn_y}. We plot the stable state $(t\to\infty)$ for each of the magnetization components and investigate the effect of varying the anisotropy strength $K$ (top row), the combined effect of exchange field and width of the ferromagnetic layer $\beta_3 \propto hL$ (middle row), and the interface barrier transparency $Z$ (bottom row). Several observations can be made. Whereas the qualitative behavior of the $m_x$ (left column) and $m_z$ (right column) components are equivalent, displaying a symmetry around $\gamma=\pi$, the $m_y$ (middle column) component displays different behavior. For some parameter values, we observe very fast oscillations in terms of the value of the stable state as a function of the superconducting phase difference. Remarkably, this is a direct result of the presence of an anomalous supercurrent in the system. To see this, consider the LLG-equation for a stable, time-independent magnetization:
\begin{align}
\textbf{m} \times \textbf{H}_\text{eff} = 0,
\end{align}
where $\textbf{H}_\text{eff}$ contains a contribution from both the anisotropy and ABS-energies. From the definition of the effective field, one can show that the components of it satisfy:
\begin{align}
\Big(\textbf{H}_\text{eff}\Big)^i \propto \sum_k C(\varepsilon_k)\frac{\partial \varepsilon_k}{\partial h_i}.
\end{align}
Now, the partial derivative of the ABS-energy depends strongly on which component of the field one considers. For instance, one finds $\frac{\partial \varepsilon_k}{\partial h_y} \propto \sin\gamma$ (odd function of the phase difference) whereas $\frac{\partial \varepsilon_k}{\partial h_z}$ is mainly determined by $\cos\gamma$ (even function of the phase difference). In turn, these properties also determine the symmetries of $\Big(\textbf{H}_\text{eff}\Big)_i$ with respect to $\gamma$. This observation is essential as it explains the qualitative behavior of the magnetization dynamics in Fig. \ref{fig:SFFFSmagdyn_y}. Let us write out the stable state condition componentwise where we explicitly separate the contribution from anisotropy and ABS-energies:
\begin{align}\label{eq:compy}
m_yH_\text{ABS}^z - m_zH_\text{ABS}^y - Km_ym_z&= 0,\notag\\
m_xH_\text{ABS}^z - m_zH_\text{ABS}^x &= 0,\notag\\
m_xH_\text{ABS}^y - m_yH_\text{ABS}^x+ Km_xm_y &= 0.
\end{align}
There are now three possible scenarios: 1) the anisotropy term dominates, 2) the ABS-energy term dominates, or 3) the contribution from both of these are comparable. When the anisotropy term dominates the effective field, one would expect that the magnetization does not deviate much from its original configuration (along the easy axis). This is seen in panel (e) for the black line. When the anisotropy term is small compared to $H_\text{ABS}$, we can neglect the terms $\propto K$ in Eq. (\ref{eq:compy}) which allows us to conclude the following: since $H_\text{ABS}^y$ is close to antisymmetric in $\gamma$ whereas $H_\text{ABS}^z$ is close to symmetric, the first and third line dictate that $m_y$ must be close to antisymmetric in $\gamma$ whereas $m_x$ and $m_z$ must be close to symmetric. This is again consistent with Fig. \ref{fig:SFFFSmagdyn_y}. Therefore, we may conclude that it is the appearance of the anomalous supercurrent (which is proportional to the $\sin\gamma$ term in the effective field) that is responsible for the qualitatively different behavior of $m_y$ compared to the other components. Finally, the oscillatory behavior of $m_y$ may be understood as a competition between the anisotropy and the ABS-contribution to the effective field. Whereas dominating $K$ permits a symmetric $m_y$ with respect to the phase difference $\gamma$ while dominating ABS-contribution gives an antisymmetric $m_y$, the two terms compete when they are of comparable magnitude and give rise to a stable-state for $m_y$ which displays symmetry in a certain range of $\gamma$ and otherwise antisymmetry. Having established the influence of the superconducting phase difference on the magnetization dynamics, the plots moreover show that magnetization switching is possible. For instance, panel (l) shows that depending on the phase difference $\gamma$, the stable magnetization state is almost fully aligned with either the $+\hat{z}$ or the $-\hat{z}$ direction. 

\begin{figure}[t!]
\centering
\resizebox{0.48\textwidth}{!}{
\includegraphics{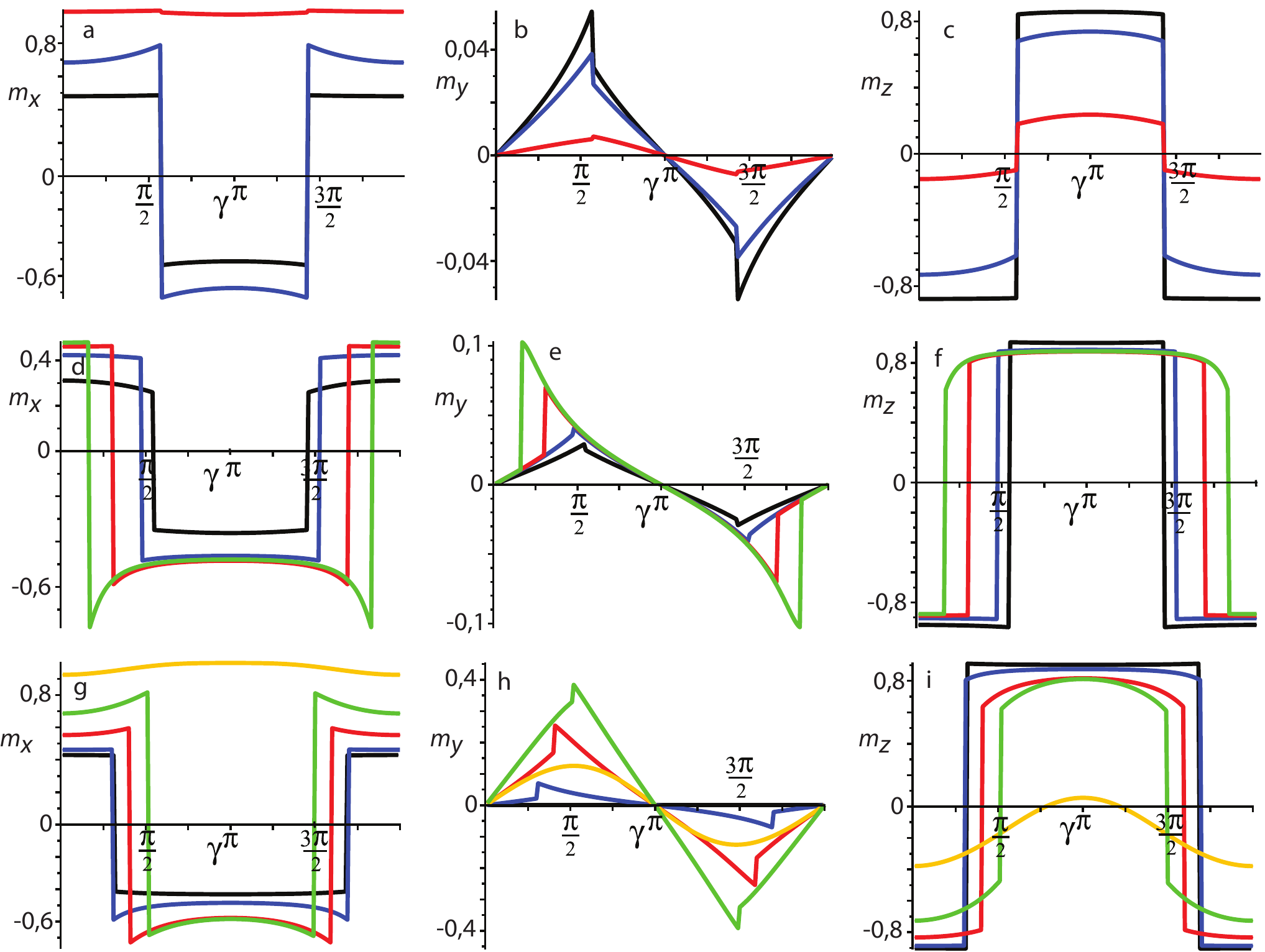}}
\caption{(Color online) Stable magnetization state as a function of superconducting phase difference $\gamma$ for $t\to\infty$ when $\textbf{m}_3 \parallel \hat{y}$ initially. The components of the magnetization are given in the left ($m_x$), middle ($m_y$), and right $(m_z)$ columns. For all panels, we fix $\beta_1=\beta_2=\pi/3$. (a,b,c): $\textbf{m}_3(t=0) \parallel \hat{x}$ as initial condition with $\beta_3=\pi/100$ and $Z=0.5$. We consider several values of the anisotropy constant $K=10^3$J/m$^3$ (black line), $10^4$ (blue line) $10^5$ (red line). (d,e,f): $\textbf{m}_3(t=0) \parallel \hat{z}$ as initial condition with $Z=0.5$ and $K=10^4$J/m$^3$. We here consider different values of the $\beta_3$ parameter - $\beta_3=\pi/100$ (black), $15\pi/100$ (blue), $25 \pi/100$ (red). (g,h,i): $\textbf{m}_3(t=0) \parallel \hat{z}$ as initial condition with $\beta_3=15\pi/100$ and $K=10^4$J/m$^3$. We consider several choices for the barrier transparency $Z=0$ (black), $0.5$ (blue), $1$ (red), $1.5$ (green), $3$ (yellow). }
\label{fig:SFFFSmagdyn_xz} 
\end{figure}

Consider next the case where we change the initial magnetization configuration of the free ferromagnetic layer to be along the $\hat{x}$ or $\hat{z}$ directions. The results are shown in Fig. \ref{fig:SFFFSmagdyn_xz}. The corresponding equation governing the stable-state now changes compared to Eq. (\ref{eq:compy}) since the anisotropy contribution will now \textit{always} appear in the second line. As a result, one concludes that regardless of the strength of the anisotropy and regardless of whether the initial configuration is along $\hat{x}$ or $\hat{z}$, the $m_y$ component will always be close to antisymmetric in $\gamma$, as seen in Fig. \ref{fig:SFFFSmagdyn_xz}.

Let us also comment specifically on the role played by the interface barrier potential $Z$ and the parameter $\beta_3 \propto hL$ in terms of how they influence the magnetization dynamics. A common feature for both Fig. \ref{fig:SFFFSmagdyn_y} and \ref{fig:SFFFSmagdyn_xz} is that the $m_y$-component grows with increasing barrier $Z$. This should be seen in conjunction with that the magnitude of the anomalous supercurrent also increases with $Z$ (up to $Z\simeq 2$), as shown in Fig. \ref{fig:energy for SFFFS}. In effect, the anomalous supercurrent increases in magnitude with $Z$ and is seen to have a feedback-effect on the magnetization in terms of enhancing the magnitude of $m_y$. With regard to the role of $\beta_3$, its main role is seen to oppose the effect of the anisotropy. As $\beta_3$ increases, the influence of the ABS-contribution to the effective field becomes more dominant as evidenced by the emergent antisymmetric $m_y$ dependence on $\gamma$.

\subsection{S$\mid$F$\mid$S junction with spin-active interface zones}

We proceed to consider the structure shown in Fig. \ref{fig:model}(b): an SFS junction where the interface are spin-active. More specifically, we allow (as before) for an arbitrary magnetization direction in the free ferromagnetic layer whereas the interface regions are modeled via Eq. (\ref{eq:sa}) in the perpendicular configuration in order to allow for the possibility of spin chirality breaking with the interface moments and the bulk moment all pointing along different axes. In the quasiclassical regime of a sufficiently weak ferromagnet, we find the following analytical expression for the ABS-energy: 
\begin{align}
\varepsilon_j = \Delta_0\sqrt{1 - \mathcal{A}\cos\gamma-\mathcal{B} (h_y/h) Z^2\rho_m^2 \alpha\sin\gamma - \mathcal{C} \pm \sqrt{ \mathcal{D}(\gamma)}}
\end{align}
where the coefficients $\mathcal{A}, \mathcal{B}, \mathcal{C}$ are independent on the phase difference $\gamma$. The quantity $\mathcal{D}(\gamma)$ is a rather large expression which depends on $\gamma$; the essential property of this quantity is nevertheless that 
\begin{align}
\frac{\partial \mathcal{D}(\gamma)}{\partial \gamma}\Bigg|_{\gamma=0} \propto \mathcal{B} (h_y/h) Z^2\rho_m^2 \alpha.
\end{align}
Similarly to the trilayer structure the sin$(\gamma)$ contribution is \textit{only present} when $h_y\neq0$ and is accompanied by an anomalous supercurrent. The effect increases with the strength of the interface barrier $Z$ and its existence is actually contigent on a non-zero $Z$. Therefore, the same conclusion as for the trilayer structure holds here: chiral spin-symmetry breaking is not a sufficient criterion for the appearance of an anomalous supercurrent - it also requires scattering at the interfaces.

\begin{figure}[t!]
\centering
\resizebox{0.48\textwidth}{!}{
\includegraphics{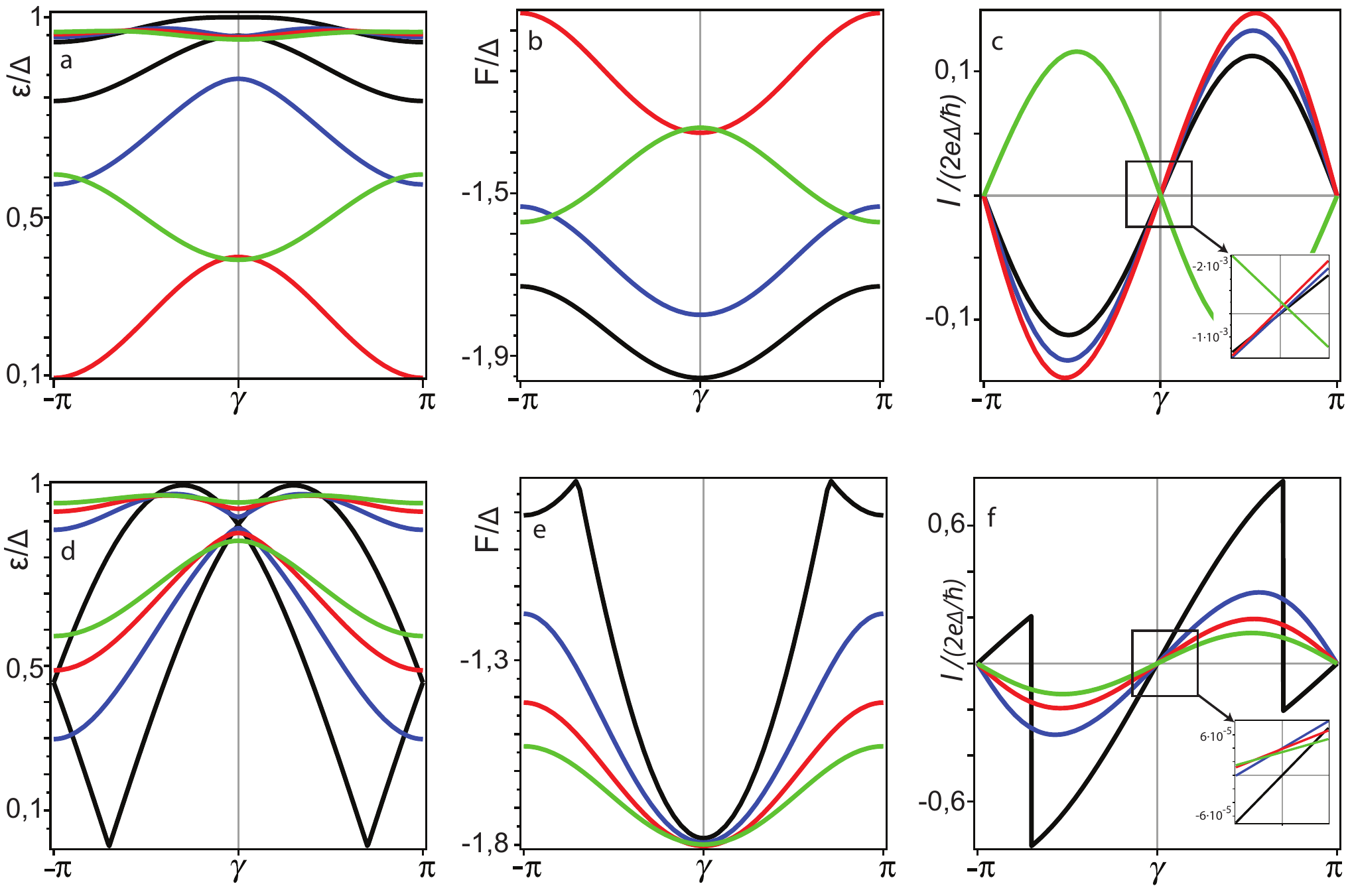}}
\caption{(Color online) (a,d): Andreev bound-state energies as a function of superconducting phase difference $\gamma$. (b,e): free energy of the system as a function of $\gamma$ and (c,f) supercurrent-phase relation for our spin-active SFS structure. In (a,b,c), we fix the barrier at $Z=2$ and investigate the effect of different values of $\beta$ (proportional to both exchange field $h$ and width $L$ of the free ferromagnetic layer): $\beta=0$ (black), $15\pi/100$ (blue), $25\pi/100$ (red), $50\pi/100$ (green). For (d,e,f), we fix $\beta=15\pi/100$ and investigate the effect of a varying barrier potential: $Z=0$ (black), $1$ (blue), $1.5$ (red), $2$ (green).}
\label{fig:energy for spin-active interface} 
\end{figure}

In Fig. \ref{fig:energy for spin-active interface}, we provide a plot for the ABS-spectrum, free energy, and supercurrent-phase relation for the system with spin-active interfaces. In this structure, there is a new parameter compared to the trilayer case, namely the ratio between the magnetic and non-magnetic part of the barrier $\rho_m$. In what follows, we set $\rho_m=0.5$. Considering first the ABS-spectrum, we see that the shift of the extremal values away from 0 and $\pi$ are very small when the conditions for a non-zero anomalous supercurrent are present (finite $Z$ and $h_y$). In fact, the free energy plots are very close to describing usual 0-$\pi$ transitions. However, the zoom-in in the right column of Fig. \ref{fig:energy for spin-active interface} demonstrates that there is a small but finite value of the supercurrent at $\gamma=0$, which is equivalent to saying that the junction is in a $\varphi$-state. Both the present and the trilayer system can then in principle act as phase batteries supplying whichever phase difference that may be desirable as its ground-state.

For the magnetization dynamics, we consider in this section only the case where the initial configuration is along the $\hat{y}$-axis since this gives the qualitatively most interesting behavior. Using the $\hat{x}$ and $\hat{z}$ directions as the free layer initial state provides similar results as in the previous section. One key difference is nevertheless that unlike the trilayer case, there is no magnetization dynamics whatsoever in the present scenario when $Z=0$. The reason is that for perfectly transparent interfaces, the junction is equivalent to a homogeneous SFS junction and there is no spin-transfer torque due to misaligned magnetic moments. Moreover, we see that for all parameter choices we have $m_x(t\to\infty) = m_z(t\to\infty)$. This stems from the fact that the influence of both spin-active interfaces is equivalent in magnitude so that the induced $x$ and $z$-components of the bulk magnetization take the same values. The qualitative behavior of the stable-state magnetization $m_y(t\to\infty)$ is determined by the relative contribution of the anisotropy term and the ABS-energies, and a similar analysis as for the trilayer case holds here as well. With increasing $\beta \propto hL$, the influence of the anisotropy term decreases.

\begin{figure}[t!]
\centering
\resizebox{0.50\textwidth}{!}{
\includegraphics{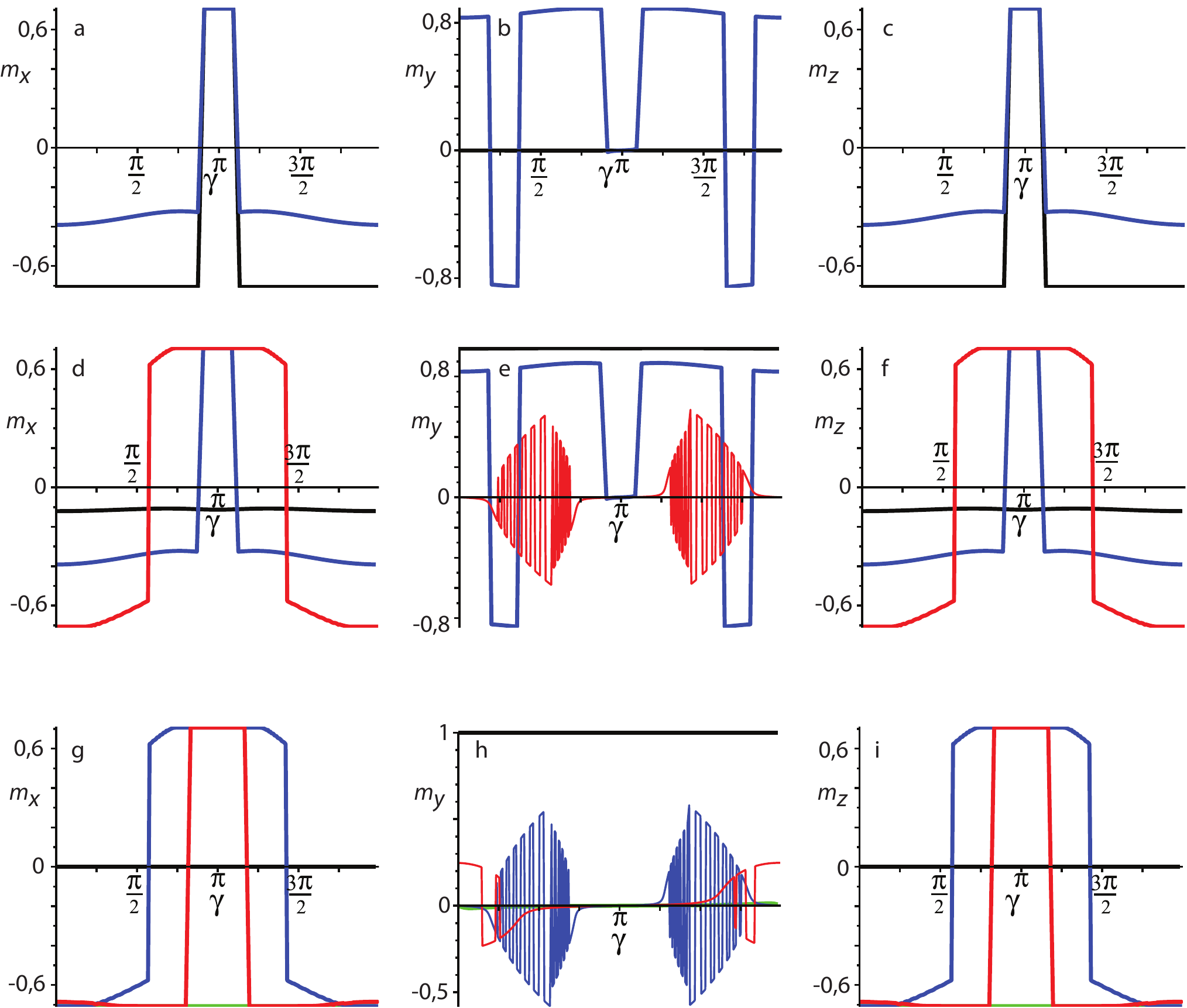}}
\caption{(Color online) Stable magnetization state as a function of superconducting phase difference $\gamma$ for $t\to\infty$ when $\textbf{m}_3(t=0) \parallel \hat{y}$ initially. The components of the magnetization are given in the left ($m_x$), middle ($m_y$), and right $(m_z)$ columns. In all panels, we fix $\rho_m=0.5$. (a,b,c): We fix $\beta=15\pi/100$, $Z=0.5$, and consider several values of the anisotropy constant - $K=10^4$ J/m$^3$ (black line), $10^5$ (blue line). (d,e,f): We fix $Z=0.5$, $K=10^5$, and consider several values of the $\beta$ parameter - $\beta=5\pi/100$ (black), $15\pi/100$ (blue), $25 \pi/100$ (red). (g,h,i): we fix $\beta=25\pi/100$, $K=10^5$, and consider several values of the interface transparency - $Z=0$ (black), $0.5$ (blue), $1$ (red), $2$ (green).}
\label{fig:magnetization for spin-active} 
\end{figure}

\subsection{Domain wall S$\mid$F$\mid$S junction}
The final structure under consideration in this work is one where the magnetic weak link connecting the superconductors consists of two layers: a magnetic domain wall ferromagnet and, as before, a free ferromagnetic layer. The domain wall is modeled via Eq. (\ref{eq:dwprofile}). In the quasiclassical regime $h\ll\mu$, we obtain the expression
\begin{align}\label{eq:absdw}
\varepsilon_j = \Delta_0\sqrt{1 - \mathcal{A}\cos\gamma-\mathcal{B}\pm \sqrt{(\mathcal{A}\cos\gamma )^2 + \mathcal{C}\cos\gamma+\mathcal{D}}}
\end{align}
where all coefficients A, B, C and D are independent of $\gamma$ and instead depend on all the other parameters in the junction. In obtaining Eq. (\ref{eq:absdw}), we considered the limit $\eta\ll1$ and $\alpha_\text{dw} \gg \eta$ where 
\begin{align}
\alpha_\text{dw} &= h_\text{dw}/2\mu,\notag\\
\eta &= a^2/k_F^2,\; a = \pi/2l_\text{dw}.
\end{align}

\begin{figure}[t!]
\centering
\resizebox{0.50\textwidth}{!}{
\includegraphics{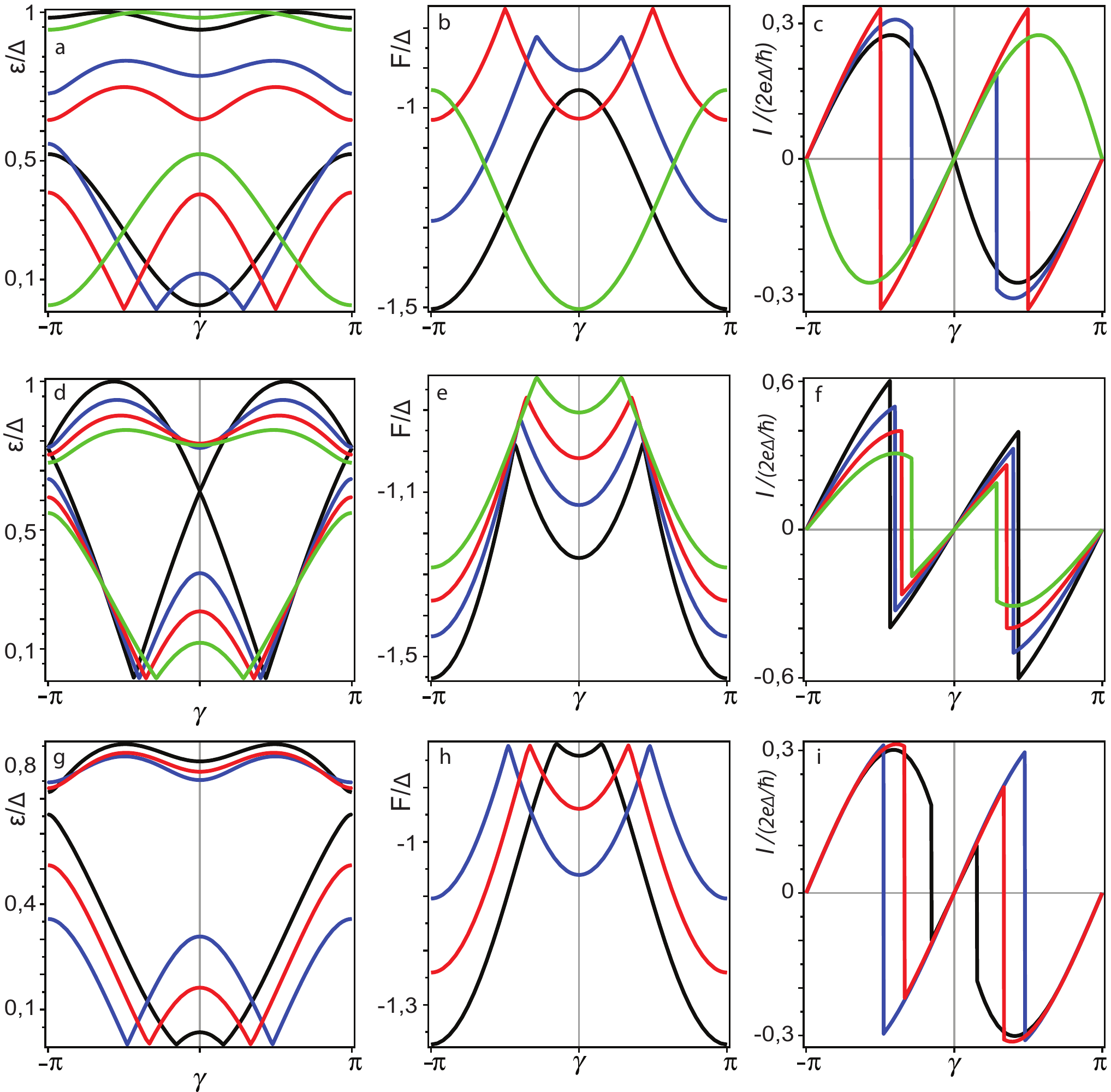}}
\caption{ (Color online) (a,d,g): Andreev bound-state energies as a function of superconducting phase difference $\gamma$. (b,e,h): free energy of the system as a function of $\gamma$ and (c,f,j): supercurrent-phase relation for our S/DW/F/S structure. In (a,b,c), we set $Z=2$, $\eta=10^{-4}$ and investigate the effect of different values of the $\beta_2$ parameter - $\beta_2=0$ (black), $15\pi/100$ (blue), $25\pi/100$ (red), $50\pi/100$ (green). In (d,e,f), we set $\beta_2=15\pi/100$, $\eta=10^{-4}$ and investigate the effect of the magnitude of the barrier transparency - $Z=0$ (black), $1$ (blue), $1.5$ (red), $2$ (green). Finally, for (g,h,i) we set $Z=2$, $\beta_2=15\pi/100$ and investigate the effect of the domain wall width - $\eta=10^{-4}$ (black), $8\times10^{-3}$ (blue), $5\times10^{-3}$ (red) }
\label{fig:energies for domainwall} 
\end{figure}

To understand what this limit means physically, we note that it is equivalent to stating that the domain wall width $\l_\text{dw}$ far exceeds a typical lattice spacing constant as it should. From this expression, it is clear that the ground-state energy will always occur at $\gamma=0$ or $\gamma=\pi$, in contrast to the two previously analyzed configurations. The $\sin\gamma$ term responsible for the anomalous supercurrent and $\varphi$-junction is absent. For this reason, we do not include any results for the magnetization dynamics of this system. We instead show graphically in Fig. \ref{fig:energies for domainwall} the ABS-energies (a,d,g), the free energy of the system (b,e,h) and the supercurrent-phase relation (c,f,j) are all shown for various parameter choices. The fact that the anomalous supercurrent is absent is an important observation, because it demonstrates that chiral spin-symmetry breaking (or alternatively, non-coplanar magnetization vectors) alone is insufficient to induce such a term. In fact, the finding that the term causing a $\varphi$-junction is absent in the present case of a domain wall is consistent with our findings for the trilayer junction above. There, it was shown that if either interface barrier between the ferromagnetic layers was absent, the anomalous supercurrent vanishes. Such a scenario is similar to the present case, since two misaligned ferromagnetic regions without any interface scattering barrier can be thought of as a simplified domain wall.

\section{Discussion}\label{sec:discussion}

We discuss here some issues which are relevant for the approximations made in our model as well as how to realize experimentally the proposed setups. First of all, the variation of the magnetization dynamics on the superconducting phase difference can be probed in several ways. In our treatment, we have considered a phase-biased Josephson junction with a fixed superconducting phase rather than a fixed current bias. In the latter case, the superconducting phase would vary together with the magnetization dynamics since the supercurrent-phase relation is sensitive to the exact magnetization configuration. Instead, by phase-biasing the junction via a loop-geometry and a minute external field corresponding to a flux quantum (which has no effect on the magnetization dynamics), the current is allowed to vary as the magnetization dynamics takes place while the phase remains fixed. Another approach would be to study a phase-driven junction with a voltage-bias as done in \eg \cite{konschelle_prl_09, linder_prb_11}. 

For the computation of the magnetization dynamics, we used as initial condition that the magnetization of the free layer was along the easy axis anisotropy. In general, however, the magnetization configuration that solves the static LLG equation in equilibrium is not necessarily with the free layer along the easy axis. This is due to the presence of the effective field stemming from the ABS-energies that exist in the junction. We have attempted to find a general analytical solution for the orientation of the free layer which solves $\mathbf{m}\times\mathbf{H}_\text{eff}=0$ when including all terms in the free energy, but the resulting expressions were too cumbersome to be of any use. The initial condition used in the numerical simulations is nevertheless feasible to realize experimentally, simply by applying an external field along the anisotropy axis to artificially enhance it so that the free layer $\mathbf{m}$ is fixed along that direction. By then turning off the field, the resulting magnetization due to the Andreev-bound states and the change in superconducting phase difference may then be observed. It is important to underline that the supercurrent-induced magnetization dynamics studied in this paper is a non-equilibrium effect even when the SC phase difference is kept constant. The reason is that the system is initially prepared in a magnetization configuration which is not the ground-state of the system so that there is a finite torque acting on the free layer which eventually goes to zero as the system relaxes into a stable state for $t\to\infty$. 

In the situation considered in the majority of previous literature on magnetic Josephson junctions, the magnetization is considered fixed and thus already being in its ground-state (\eg due to strong anisotropy fixing). One then assumes that there is no feedback on the magnetization from the Josephson current, and so one only needs to minimize the superconducting part of the free energy with respect 
to the phase difference: the magnetic part is already assumed to be minimized. If one instead, as we have done, allows for the Andreev bound states and (thus supercurrent) to have a considerable influence on the free energy on equal footing as the anisotropy, the superconducting correlations will alter 
the favorable orientation of the magnetization. The free energy should then be minimized both with respect to the magnetization orientation and the superconducting phase difference.

Let us also comment specifically on our technical treatment of how the Andreev-bound state contribution to the free energy gives rise to an effective field that enters the LLG equation. By defining the effective field $\mathbf{H}_\text{eff}$ as the functional derivative of the magnetic order parameter evaluated at its instantaneous configuration requires that the magnetization dynamics is slow compared to relaxation processes in the system. In other words, the derived free energy may be treated as time-dependent if the system approximately equilibriates in pace with the change in magnetization. A lag between the magnetization dynamics $\mathbf{m}(t)$ and degrees of freedom that are coupled to it may be interpreted as a dissipation of energy and in turn captured by the Gilbert-damping parameter that we have accounted for \cite{tserkovnyak_rmp_05}. For a driven superconducting phase where the phase difference is $\gamma(t) = \omega_Jt + \gamma_0$, the above criterium is satisfied when $\omega_J\ll k_BT_c$ \cite{konschelle_prl_09} so that the phase is treated as a time-dependent perturbation.

In order for the magnetization vectors to be misaligned as \eg in the trilayer case, it is necessary to reduce the exchange coupling between the layers. This can be achieved by inserting a normal metal spacer between the F regions. We have omitted this layer in our calculations since it would merely complicate the analytical expressions without introducing any new physics. It should be noted that spacer thicknesses as small as 4 nm are sufficient to experimentally allow for misaligned magnetization vectors in superconducting hybrid structures, as very recently reported in \cite{jara_arxiv_14}.

\section{Conclusion}\label{sec:summary}
In conclusion, we have investigated the spin- and charge-transport in several models of magnetically textured Josephson junctions. We have made predictions for the ABS-energy spectrum, the free energy and its phase dependence, and the supercurrent-phase relation. Moreover, we have considered the magnetization dynamics induced by the presence of a triplet spin-supercurrent in these systems and computed how the stable-state magnetization $\mathbf{m}(t\to\infty)$ is controlled by the superconducting phase difference. A key finding is that the presence of an anomalous supercurrent $\propto \cos\gamma$, which results in a $\varphi$-state, strongly influences the resulting magnetization dynamics and gives rise to symmetry properties of the stable-state which may be understood by analyzing the resulting effective field $\mathbf{H}_\text{eff}$. Moreover, we demonstrated that chiral spin symmetry breaking is insufficient to generate such an anomalous supercurrent: the presence of scattering barriers separating different magnetic regions play an instrumental role in creating this effect. Our results may provide a basis for future investigations of how controllable magnetization dynamics can be obtained with spin-supercurrents that are tuned via the superconducting phase difference.

\acknowledgments

The authors thank J. W. A. Robinson, K. Halterman, and D. Kulagin for very useful discussions. I.K and J.L were supported by the Research Council of Norway, Grant No. 205591/F20 (FRINAT).

\appendix

\section{Calculation of Andreev levels}\label{sec:appendix}

In order to solve the Bogoliubov-de Gennes equations we write the wavefunction in plane-wave form $\Psi(y)=e^{iky}\psi$. The wave vectors of electron- and hole-like quasiparticles inside the superconductor are:
\begin{equation}
k_S = \sqrt{2m(\mu \pm \sqrt{E^2-\Delta^2})}
\end{equation}
while for the homogeneous ferromagnets we have:
\begin{equation}
k_f^\sigma=\sqrt{2m(\mu \pm E+\sigma h)}.
\end{equation}
Finally, in the domain wall case we find:
\begin{equation}
k_{DW}^\sigma=\sqrt{2m(\mu \pm E)+a^2 + \sigma 2\sqrt{2m a^2 (\mu\pm E)+m^2 h^2}}
\end{equation}
where $a=\frac{\pi}{2 l_{DW}}$. Defining $\alpha=\frac{h}{2\mu}$ and $\eta=\frac{a^2}{2m(\mu \pm E)}$, we find in the limit $E\ll \mu$ that $\eta=\frac{a^2}{k_F^2}$  and the wave vector for the domain wall becomes:
\begin{equation}
k_{DW}^\sigma=\sqrt{2m(\mu \pm E)+\eta^2 + \sigma 2\sqrt{\eta^2+\alpha^2}}
\end{equation}
During our calculation we use the approximation that $E\ll\mu$ and that $\alpha$ and $\eta$ are small. For $\alpha\gg\eta$, the wavevector for the quasiparticles in the domain wall ferromagnet can be simplified further:
\begin{equation}
k_{DW}^\sigma= 2m\mu (1+\sigma \alpha)
\end{equation}
while for for $\alpha\ll\eta$
\begin{equation}
k_{DW}^\sigma= 2m \mu (1+\sigma \eta)
\end{equation}

\begin{widetext}
For a ferromagnetic layer with arbitrary orientation of magnetization, we have:
\begin{equation}
\Psi_F(y)=\sum\limits_{p=\pm}  \Biggl( t_{e, \uparrow}^{\pm} \begin{pmatrix}
\cos(\frac{\theta}{2})\\
\sin(\frac{\theta}{2}) e^{i\chi}\\
0\\
0
\end{pmatrix} e^{\pm ik_F^\uparrow y}+ t_{e, \downarrow}^{\pm} \begin{pmatrix}
-\sin(\frac{\theta}{2}) e^{-i\chi}\\
\cos(\frac{\theta}{2})\\
0\\
0
\end{pmatrix} e^{\pm ik_F^\downarrow y}+t_{h,\uparrow}^{\pm} \begin{pmatrix}
0\\
0\\
\cos(\frac{\theta}{2})\\
\sin(\frac{\theta}{2}) e^{-i \chi}
\end{pmatrix} e^{\pm (-ik_F^\uparrow y)}+t_{h, \downarrow}^{\pm} \begin{pmatrix}
0\\
0\\
-\sin(\frac{\theta}{2}) e^{i \chi}\\
\cos(\frac{\theta}{2})
\end{pmatrix} e^{\pm (-ik_F^\downarrow y)} \Biggr)
\end{equation}
where $\theta$ is the angle between the magnetization and the $z$-axis, $\chi$ is the angle between the magnetization and the $x$-axis in the $x-y$-plane, $\pm$ corresponds to the direction of the moving particles. For the domain wall layer, we first perform a unitary transformation $\hat{\mathcal{U}}$ of the Hamiltonian to remove the explicit spatial dependence of the exchange field due to the domain wall texture. This is achieved by rotating the system so that the local spin quantization axis is aligned with the local magnetization direction. Starting out with $\hat{H}\psi = \varepsilon\psi$, we rewrite it to $\hat{H}_\text{rot}\Psi = \varepsilon\Psi$ where $\hat{H}_\text{rot} = \hat{\mathcal{U}}\hat{H}\hat{\mathcal{U}}^{-1}$ and $\Psi = \hat{\mathcal{U}}\psi$. The new wavefunction $\Psi$ may then be expressed as follows:
\begin{equation}
\Psi_{DW}(y)=\sum\limits_{p=\pm}  \Biggl( t_{e, \uparrow}^{\pm} \begin{pmatrix}
\phi_1^\uparrow\\
\pm \phi_2^\uparrow\\
0\\
0
\end{pmatrix} e^{\pm ik_{DW}^\uparrow y}+ t_{e, \downarrow}^{\pm} \begin{pmatrix}
\pm\phi_2^\downarrow\\
\phi_1^\downarrow\\
0\\
0
\end{pmatrix} e^{\pm ik_{DW}^\downarrow y}+t_{h,\uparrow}^{\pm} \begin{pmatrix}
0\\
0\\
\phi_1^\uparrow\\
\pm \phi_2^\uparrow
\end{pmatrix} e^{\pm (-ik_{DW}^\uparrow y)}+t_{h, \downarrow}^{\pm} \begin{pmatrix}
0\\
0\\
\pm \phi_2^\downarrow\\
\phi_1^\downarrow
\end{pmatrix} e^{\pm (-ik_{DW}^\downarrow y)} \Biggr)
\end{equation}

where  \begin{equation}
\phi_1^\sigma=\sigma(\alpha+\eta^2 \sqrt{\alpha^2+\sigma \eta^2}), 
\phi_2^\sigma=\sigma i\eta\sqrt{1+\eta^2+2\sqrt{\alpha^2+\eta^2}} 
\end{equation}
\end{widetext}
We may then revert to the original wavefunction $\psi$, which enters the boundary conditions, by doing the inverse transformation $\psi = \hat{U}^{-1}\Psi$. The coefficients $t_{e(h), \sigma}^{\pm}$ are associated with right- left-going $(\pm)$ ELQ and HLQ propagating throught the ferromagnetic layers. The spin index $\sigma= \uparrow$ or $\downarrow$.


The wave functions must satisfy the boundary conditions of 1) continuity of the wave function at the boundary: 
\begin{equation}
(\Psi_k-\Psi_l)|_{y=L_i}=0
\end{equation}
and 2) discontinuity of the first derivative at the boundary:
\begin{equation}
\partial(\Psi_k-\Psi_l)|_{y=L_i}=\frac{2m}{\hbar}U \Psi|_{y=L_i}
\end{equation}
where $L_i=0$, $L_1$, $L_2$, $L_3$, indexes $k$ and $l$ are associated with corresponding index of the wave functions. We have defined the normalized barrier strength $Z = 2mU/(\hbar k_F)$. Note that in the domain wall case, extra terms $\partial_y\hat{\mathcal{U}}$ arise in the boundary conditions due to the unitary transformation of the wavefunction. From the boundary conditions, we can obtain all the scattering coefficients and set up a homogeneous system of linear equations, demanding that the determinant is equal to zero in order to have a non-trivial solution. The resulting characteristic equation is then solved for the energy which represents the Andreev Bound State (ABS). In the ABS-energy for the trilayered structure, the coefficient $B$ before the anomalous $\sin\gamma$ term satisfies
\begin{equation}
B \propto \sin{2\beta_1} \sin{2\beta_2} \sin{2\beta_3}
\end{equation}
whereas for the structure with spin-active interfaces
\begin{equation}
B \propto \sin^2 {\beta}.
\end{equation}
In the scenario with a domain wall ferromagnet, there exists is no managable expression for $B$ in the general case.

\end{document}